\documentstyle[prl,epsfig,aps]{revtex}

\def \bi{\bibitem}

 \def\(({\left(}
 \def\)){\right)}
\def\bi{\bibitem}

\def \ov{\over}

\def \a{\alpha}
\def \b{\beta}

\def \z{\zeta}

\def \e{{\rm e}}
\def \del{\delta}

\def \beqna{\begin{eqnarray}}
\def \eeqna{\end{eqnarray}}
\def \beq{\begin{equation}}
\def \eeq{\end{equation}}

\def \ov{\over}

\def \a{\alpha}

\def \b{\beta}

\def \ra{\rangle}
\def \ab2{\alpha\beta^2}
\def \ra{\rangle}

 \newcommand \be {\begin{equation}}
\newcommand \bea {\begin{eqnarray} \nonumber }
\newcommand \ee {\end{equation}}
\newcommand \eea {\end{eqnarray}}

\newcommand \la {\langle}

\input{epsf}
\begin{document}
\twocolumn[\hsize\textwidth\columnwidth\hsize\csname@twocolumnfalse\endcsname
\preprint{MA/UC3M/11/95}
\title{Relaxation processes and entropic traps in the Backgammon model}

\author{ Silvio Franz(*) and Felix Ritort(**)}
\address{
(*) International Center for Theoretical Physics\\
Strada Costiera 11\\
P.O. Box 563\\
34100 Trieste (Italy)\\
(**) Institute of Theoretical Physics\\
University of Amsterdam\\
Valckenierstraat 65\\
1018 XE Amsterdam (The Netherlands)\\
e-mail: {\it franz@ictp.trieste.it }\\
e-mail: {\it ritort@phys.uva.nl}}
\date{March 1996}
\maketitle

\begin{abstract}
We examine the density-density correlation function in a model recently
proposed to study the effect of entropy barriers in glassy dynamics. We
find that the relaxation proceeds in two steps with a fast beta process
followed by alpha relaxation. The results are physically interpreted in
the context of an adiabatic approximation which allows to separate the
two processes, and to define an effective temperature in the
off-equilibrium dynamics of the model. We investigate the behavior of
the response function associated to the density, and find violations of
the fluctuation dissipation theorem.

\end{abstract}
\twocolumn
\vskip.5pc] 
\narrowtext 
The relaxation in supercooled liquids near the glass transition has a
characteristic two step form. Experiments on very different materials
reveal the existence of a first fast relaxation process, called beta
relaxation, followed by a much slower one, called alpha \cite{books}.
One of the most striking successes of the Mode Coupling Theory
\cite{Gotze} is its ability to capture this phenomenon and to give a
correct prediction for the relation  
among the exponents characterizing the
two relaxations. However, we believe that a comprehension of the basic
mechanisms underlying the relaxation in glasses is missing. Experiments
in glasses have been recently interpreted in terms of traps models
\cite{odagaki,Bouchaud}.  In these models, the system evolves among
traps -or metastable states- which have a life time that grows
decreasing the temperature, and finally diverges at the glass
transition. The two step relaxation follows naturally from the
hypothesis that equilibration inside a trap occurs much faster than
``jumps'' among different traps.  How a trap may be defined and
described for real systems or microscopic models is an interesting open
problem.  If the traps have to be interpreted as the result of energy
barriers in a rough energy landscape, one finds the difficulty that the
relaxation should appear as a random process even on a large scale. A
``jump'' among two different traps should imply a discontinuity in
various quantities as the energy or the correlation function.  This
problem was already noted in \cite{Bouchaud}, where it was proposed, as
a way out, that real systems could be composed by a large number of
quasi-independent sub-systems leading to the observed self averaging
properties for the different quantities.  In this direction, it can be
instructive to investigate a different mechanism for slow
relaxation, and in particular the role of entropy barriers.

In this letter we study the nature of density fluctuations in the so called
``Backgammon model'', a microscopic model that has proved useful in
studying  some mechanisms underlying the glassy relaxation, and in
particular the role of entropy barriers. The model \cite{I} is a
Boltzmann gas with $N$ particles in an $N$ site space with Hamiltonian
given by the total number of empty sites,

\be 
H=-\sum_{r=1}^N\del_{n_r,0} 
\ee 

where $r=1,...,N$ denotes the sites of the space and $n_r$ the
occupation number of the site $r$. The system evolves following a single
particle Metropolis dynamics: at each sweep a particle to move and an
arrival site are chosen at random.  The particle is moved with
probability 1 if the energy does not increase, and with probability
$\exp(-\beta)$ if the move costs one unity of energy.  The relaxation of
the energy has been studied in detail in \cite{I,II,III,GBM,GL}, we
briefly reassume here the main results.  At zero temperature the
dynamics is slower and slower
 as the times goes by: the average density of particles
in the occupied states increases, and as a consequence the dynamics
slows down. This observation has allowed for the identification of fast
and slow degrees of freedom: the relaxation within the occupied sites at
a given time proceeds much faster then the variation of the energy and
the diffusion of ``towers of particles''.
This allows a self-consistent
treatment of the dynamics which is in very good agreement with the exact
solution. At large time the occupation probability $P(n,t)$
mimics the equilibrium one 
\be
P(n,t)=\e^{\b(t)\del_{n,0}-\z(t)}{\zeta(t)^{n-1}\ov  n!}
\ee
with effective time-dependent temperature $T(t)=1/\b(t)$
larger than the effective one, and effective fugacity $\zeta(t)$
related to $\b(t)$ by the condition of constant density $\la n_r(t)\ra=1$,
i.e. $ \e^{\b(t)}+\e^{\zeta(t)}-1=\zeta(t)\e^{\zeta(t)}$.
At zero temperature, the decay of the energy follows the law:
$E(t)\sim -1+O(1/\log(t))$ while at small but finite temperatures this behavior
is cutted-off for times of the order of the relaxation time $\tau \sim
exp(\beta)/\b^2$ when the relaxation becomes exponential.  Probing the
system on finite time scales, e.g. mimicking heating cooling
experiments, one finds characteristic glassy behavior, as hysteresis
loops for the energy \cite{II}. Additional information on the
off-equilibrium dynamics is gained studying the energy-energy
autocorrelation function $C_E(t,s)=\frac{<
\del_{n_r(t),0}\del_{n_r(s),0}>-E(t)E(s)}{E(s)(1-E(s)) }$ \cite{I}. 
This quantity shows
aging at zero temperature with a scaling behavior 
$C_E(t,s)\sim (s/t)^{\frac{1}{2}}(\log
(s)/\log (t))$ \cite{GL}. Again, at finite
temperature this scaling is observed up to times of the order of the
relaxation time.
 
Here we concentrate our attention to the relaxation of the
density-density correlation function, a quantity that is measured in
experiments on real systems \cite{books}. This is a better quantity 
to study
fast processes in the system since on short time scales 
local densities varies while energy stays essentially constant.

The fluctuations of the local density, around its average 
$\la n_r(t)\ra=1$, are studied introducing the 
 density-density 
correlation function
\begin{eqnarray}
C^{(r)}(t,s)= \la n_r(t) n_r(s) \ra=\nonumber\\
\sum_{n,m} n\; m\; P^{(r)}
(n,t |m,s)\; P^{(r)}(m,s)
\label{eqc}
\end{eqnarray}
and its associated response
\be
R^{(r)}(t,s)={\del \la n_r(t) \ra \ov \del h_r(s)}
\label{eqr}
\ee
having denoted as
 $P^{(r)}(n,t)$ the probability that state $r$ is occupied by 
$n$ particles at time $t$, $P^{(r)}(n,t |m,s)$ the same
 probability conditioned to have $m$ particles at time $s$, and 
$h_r$ is an infinitesimal inhomogeneous
 ``pressure field'' coupled linearly 
with the local density in the Hamiltonian \cite{citacio}.
We will present data for the normalized correlation 
$C_{norm}(t,s)=\frac{\la n_r(t)n_r(s)\ra-1}
{\la n_r(s)^2\ra-1}$.
Choosing a site independent initial distribution, the correlation function
remains site independent, while the response 
at finite time depends on the distribution on the 
$h_{r'}$ on the different sites, in the form of a dependence  on 
$\mu^{(r)}=Prob(h_{r'}>h_r)$ and $\nu^{(r)}=Prob(h_{r'}<h_r)$. 
Despite that, in the following we will drop the index $r$ from 
the various quantities. 
At equilibrium  correlation and response are time translation invariant
and related by the fluctuation-dissipation theorem relation 
$T R(t-s)=\partial C(t-s)/\partial s$. 
In off-equilibrium conditions it has been often
proved useful to characterize the violation of equilibrium by
the ``fluctuation dissipation ratio'' \cite{cuku,fm}
\be 
x(t,s)={T R(t,s)\ov {\partial C(t,s)\ov \partial s}}.
\ee

The evolution of the previously defined functions can be studied
starting from the hierarchy of equations for $P(n,t)$ and $P(n,t|m,s)$
and related quantities following a procedure similar to the one used
in \cite{III}. One can then derive closed integral equations in terms
of few functions that can be integrated numerically \cite{III} or
analytically in the long time limit \cite{GL}. Details of this
analysis, and in particular the full set of hierarchies for the
density-density correlation (\ref{eqc}) and the associated response
function (\ref{eqr}) will be presented elsewhere.  In this paper we
integrate directly, truncating the hierarchy to some large order.  In
practice we have found that for low enough temperature and not too
large times the truncation at $n=100$ yields excellent results.

Let us now discuss the form of the density correlation function in
equilibrium. Starting from a random initial configuration, after times
of the order of $\tau_\a\approx \e^\b/\b^2$ the system eventually
reaches equilibrium. The correlation function is then time translation
invariant and can be studied exactly in Laplace transform. It turns out that
for 
temperatures small enough
the dominant contribution to the equilibrium correlation
function for all times is given by the superposition of two Debye
processes 
\be C_{norm}^{eq}(t)= {\z-1+\e^{-t/\z}\ov \z}\e^{-t/\tau_\a}
\label{ceq}
\ee
with $\tau_\a=2((\z-1)\e^\z+1)/\z^2\approx 2 \e^\b/\b^2>>\z$.
The fast and slow
 processes at equilibrium are related respectively
to the relaxation within 
occupied states and to diffusion. We notice that 
the value of the plateau at equilibrium is $(\z-1)/\z$, that corresponds to 
complete decorrelation within occupied states, but no diffusion. 
The equilibrium curve for $T=0.05$ is   the dashed line in fig. 1. 
In the  time window shown in figure 1 the correlation function stays
essentially constant equal to the plateau value $(\z-1)/\z
\simeq 0.94$ decaying to zero afterwards. Here the alpha relaxation time 
is $\tau_\a\simeq 3\times 10^6$.

Regarding the off-equilibrium correlation function
we show in figure 1 the data for the $C_{norm}(t,s)$
for $T=0.05$ as a function of $t-s$ for various values of s.  Over the
time window we explore the system is far from being thermalized. 
The integration at $T=0$ on the same time window leads to
almost identical results. 
The $\a$ and $\b$ processes are well separated for large enough $s$.
In this case, the shape of the off-equilibrium 
relaxation curve can be understood 
qualitatively within the framework of the adiabatic approximation. 
The decorrelation 
time for the density among filled states is much smaller that the time 
needed to diffuse and/or change sensibly the energy; there must be then 
a time  scale such that we can approximate 
\be 
P(n,t |m,s)\approx \del_{m,0}\del_{n,0}+(1-\del_{m,0})(1-\del_{n,0}){P(n,s)
\ov 1-P_0(s)}
\ee
and consequently
\be
C_{norm}(t,s)={\z(s)-1\ov \z(s)}
\ee
For short times ($t-s<<s$), $\z(s)=-\log(P_1(s))$ does not vary too
much; it is then natural to describe the beta relaxation by the form
(\ref{ceq}) just substituting $\z$ by $\z(s)$. In this way we have
checked we correctly describe even the beginning of the alpha
relaxation. However the best combined
description of the slow $\a$ relaxation together with the $\b$
relaxation is obtained by functions of the aging form:
\be
C(t,s)\approx {\z(s)-1+\e^{-(t-s)/\z(s)} \ov \z(s)}{(1+b(s))
\ov \left(1+b(s){\sqrt{t}\log(t)\ov \sqrt{s}\log(s)}\right)}
\label{fit}
\ee
a form inspired by the one of the energy-energy correlation function found in 
\cite{GL}. 
At finite temperature, 
$b(s)$ is a crossover function to the form (\ref{ceq}).

\begin{figure}
\centerline{\epsfxsize=8cm\epsffile{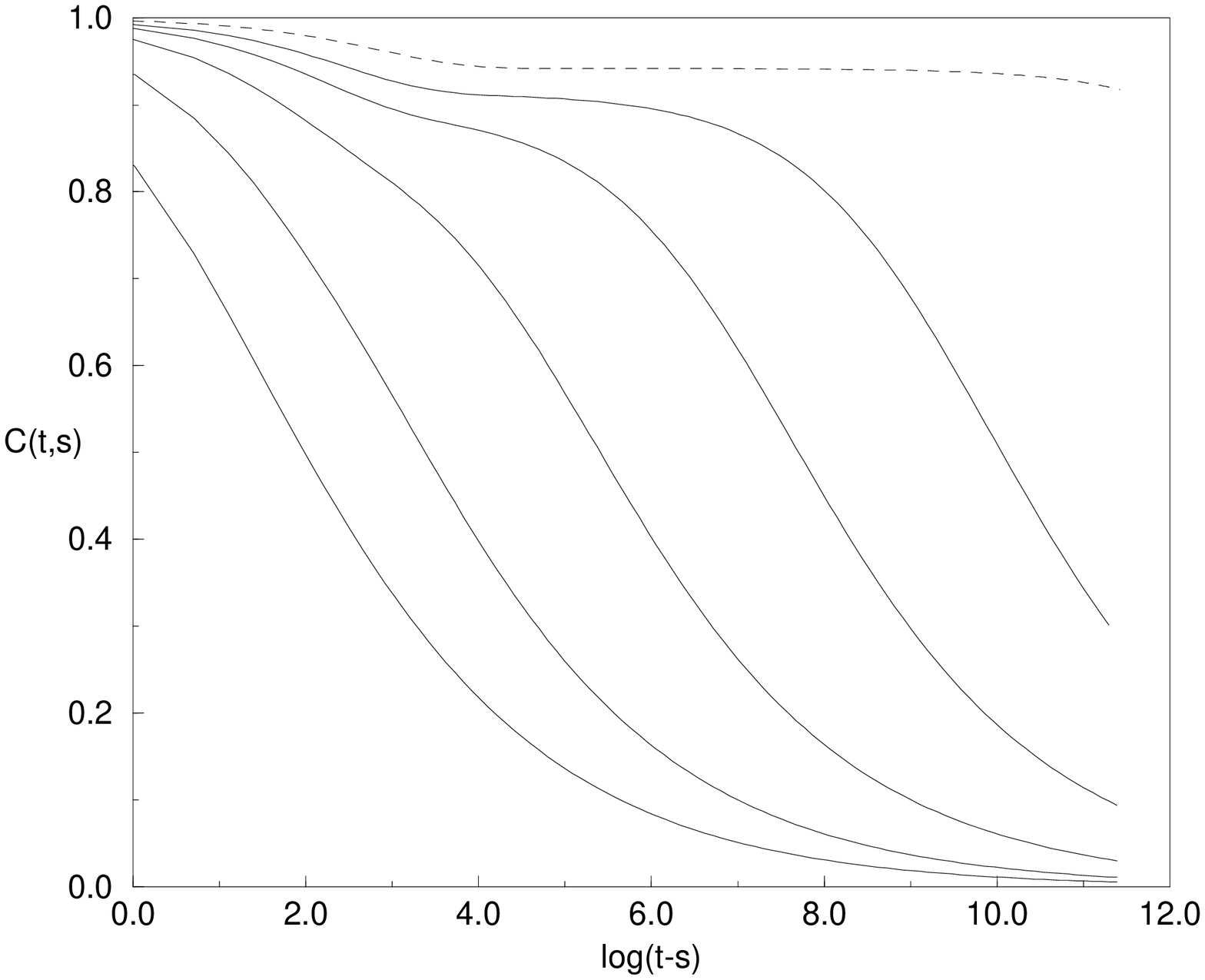}}

\caption{Density-density correlation function $C_{norm}(t,s)$ as a
function of $t-s$ for different values of $s=1,10,10^2,10^3,10^4$
(from left to right) at $T=0.05$ (continuous lines). The existence of
a plateau separates the beta and alpha regimes. The dashed line
corresponds to the equilibrium $C_{norm}^{eq}(t-s)$. The curves with
$s=10^3,10^4$ are excellently fitted by the form (\ref{fit}) with,
respectively, $b(s)=2.20,3.85$.  For the same times, $\z(s)=-\log
P(1,s)=8.93,11.51$.}

\end{figure}

In figure 2 we see that the two relaxation processes are also manifest
in the response function.  In order to characterize better the two
processes in off-equilibrium conditions we study the relation among the
response and the correlation during the dynamics.  The first quantity of
interest is the value of the fluctuation-dissipation ratio at equal
times.  We see that $x(t,t)$ reaches values close to one much before
total equilibrium sets in.  This is a further indication that the system
is in local equilibrium and yields in a natural way the notion of a trap
in this system. Note that the traps in the backgammon model are purely
entropic. This means that the system escapes from the trap even at zero
temperature when thermal excitations are absent. In other words, the
dynamics is slowed down by entropic traps which can be considered as
metastable states with a finite (energy dependent) lifetime.

\begin{figure}
\centerline{\epsfxsize=8cm\epsffile{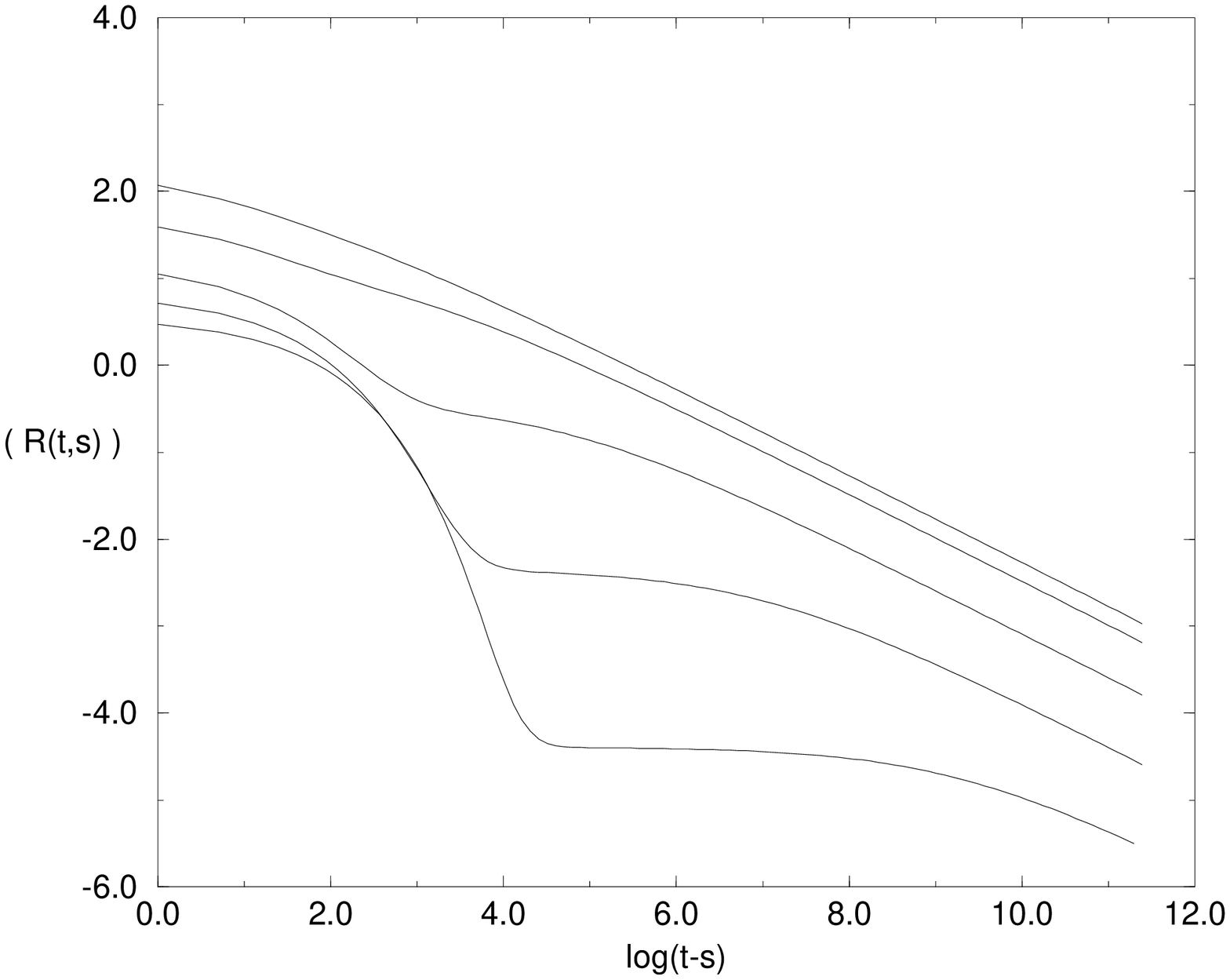}}

\caption{The response function for the values of
$s=1,10,10^2,10^3,10^4$ (from top to bottom).}

\end{figure}

\begin{figure}
\centerline{\epsfxsize=8cm\epsffile{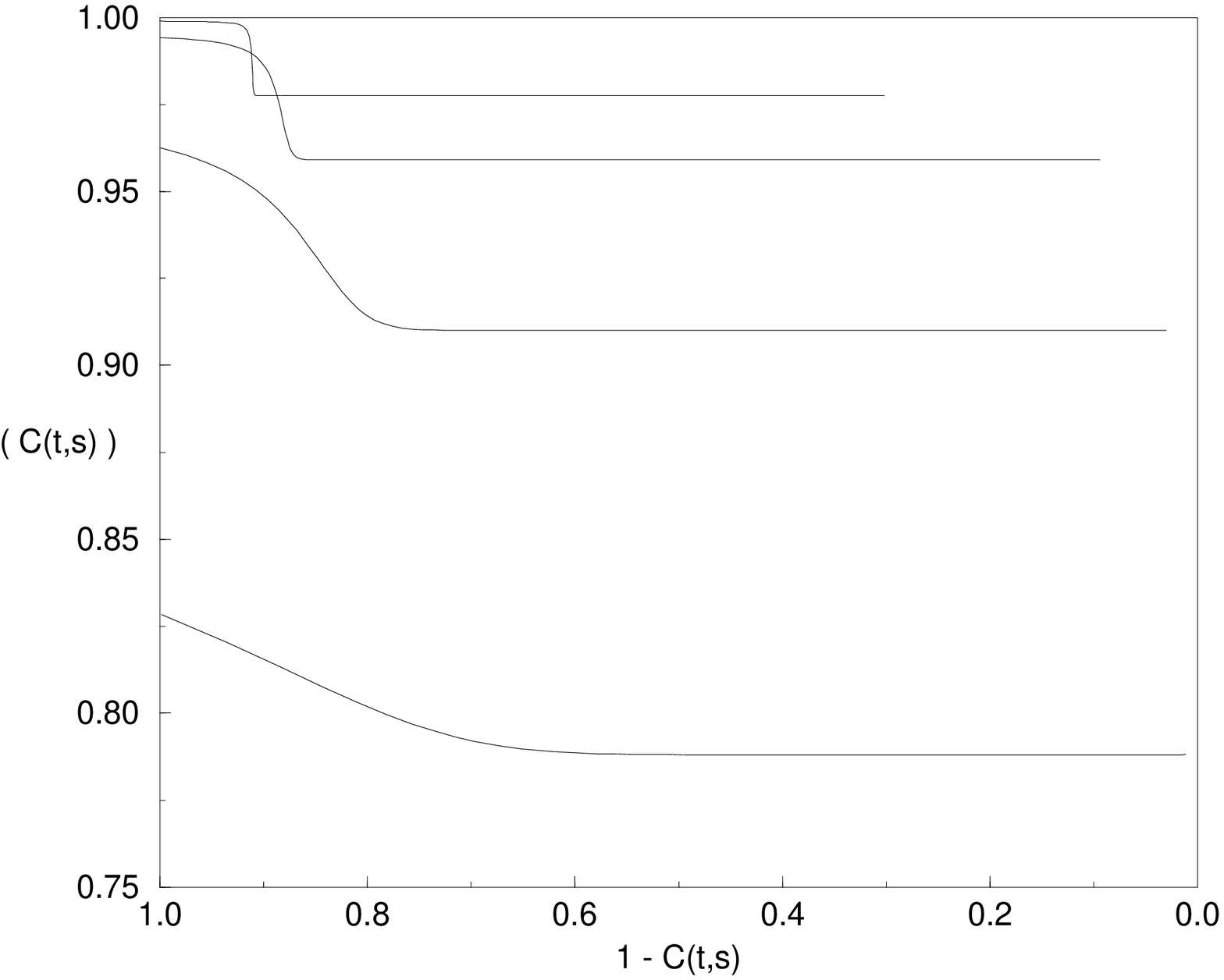}}

\caption{Fluctuation-dissipation ratio $x(t,s)$ for different values
of $s=10,10^2,10^3,10^4$ (from bottom to top) as a function of
$1-C_{norm}(t,s)$. The two plateaux correspond to the alpha and beta
relaxations.}

\end{figure}

Most interestingly we find that for large times (large values of $s$)
 $x(t,s)$ remains nearly constant in each of the two processes.
 This is shown in fig. 3 where we plot the
$x(t,s)$ as a function of $1-C_{norm}(t,s)$ for different values of $s$.
We find that the $x(t,s)$ is close to 1 in the beta relaxational process
and jumps to a $s$-dependent smaller value $x_{\alpha}(s)$ at a value
$C_{norm}(t,s)=q_{EA}$ remaining constant in the alpha process.

It is rather natural, with the adiabatic approximation in mind, 
to look for a possible interpretation of $x$ as the ratio 
between the effective temperature and the actual one. 
Unfortunately we found negative evidence for such interpretation 
of the data. For low enough temperature and times smaller than $\tau_\a$
the actual value of $x(t,s)$ is nearly temperature independent. 
 The step
behavior of the $x(t,s)$ is quite reminiscent of the
dynamics in spin
glasses with one step of replica symmetry breaking \cite{cuku}. 
This in turn represents the suitable off-equilibrium generalization 
of Mode Coupling Theory \cite{FRHE}
in case of Whitney fold glass singularity, i.e.
precisely the case where there are well separated $\a$ and $\b$ relaxations. 
We do not know at this stage if this corresponds to  a deep analogy or is
a mere coincidence. We stress that at equilibrium, the ending of the beta 
relaxation, as well as the onset of the $\a$ relaxation are described 
by power laws in Mode Coupling Theory. As it is seen in (\ref{ceq}), in our 
case the 
equilibrium relaxation
is the superposition of two exponential. 

Summarizing, we have shown the existence of two step relaxational
processes (alpha and beta relaxation) in the Backgammon model, which is
a simple model where the slow dynamics is consequence of pure entropic
barriers. We have closed the dynamical equations for the
density-density correlation function and the response of the system to
a staggered field coupled to the density. Our results for those
quantities and the fluctuation dissipation ratio allow for a clear
identification of the fast (beta) and slow (alpha) relaxation processes
in this system. The physical interpretation of this processes leads
naturally to the concept of entropic trap. In case of entropic traps,
the entropy barrier associated to the trap depends itself on the number
of available configurations within the trap. Because the height of the
entropic barriers varies itself in a continuous way, the dynamics
proceeds slow but without macroscopic jumps in the energy. The results
presented here for the Backgammon model are expected to apply for slowly
relaxing systems where the dynamics is mainly driven by entropy barriers
(like for instance, Bose-Einstein condensation). A detailed account of
our work will be given elsewhere.

Acknowledgments. (F.R) acknowledges FOM in Netherlands for financial
support and ICTP in Trieste for its kind ospitality in the final stages
of this work.

\end{document}